\documentclass[journal,twoside]{IEEEtran}

\ifCLASSINFOpdf
\usepackage[pdftex]{graphicx}

\else

\fi

\usepackage[cmex10]{amsmath}
\usepackage{amssymb}   
\usepackage{amsxtra}
\usepackage{amscd}
\usepackage{amsthm}
\usepackage{textcomp}
\usepackage{graphicx}  
\usepackage{color}

\usepackage{balance}

\usepackage{array}  

\usepackage{multirow}  

\usepackage{amsmath}
\usepackage{cite}

\usepackage{url}

\usepackage{color,soul} 
\soulregister\cite7
\soulregister\ref7
\soulregister\pageref7

\begin{document}
%

\title{{\huge Noise Modulation }}


%
%
%

\author{Ertugrul~Basar,~\IEEEmembership{Fellow,~IEEE}
            \vspace*{-0.35cm}
\thanks{The author is with the Communications Research and Innovation Laboratory (CoreLab), Department of Electrical and Electronics Engineering, Ko\c{c} University, Sariyer 34450, Istanbul, Turkey. (e-mail: ebasar@ku.edu.tr).}
}

\maketitle

\begin{abstract}

Instead of treating the noise as a detrimental effect, can we use it as an information carrier? In this letter, we provide the conceptual and mathematical foundations of wireless communication utilizing noise and random signals in general. Mainly, the concept of noise modulation (NoiseMod) is introduced to cover information transmission by both thermal noise and externally generated noise signals. The performance of underlying NoiseMod schemes is evaluated under both additive white Gaussian and fading channels and alternative NoiseMod designs exploiting non-coherent detection and time diversity are proposed. Extensive numerical and computer simulation results are presented to validate our designs and theoretical derivations.
  
\end{abstract}
\begin{IEEEkeywords}
Noise modulation (NoiseMod), thermal noise modulation (TherMod), bit error probability.  
\end{IEEEkeywords}

%
\IEEEpeerreviewmaketitle

\vspace*{-0.15cm}
\section{Introduction}

\begin{figure*}[!t]
	\begin{center}
		\includegraphics[width=1.6\columnwidth]{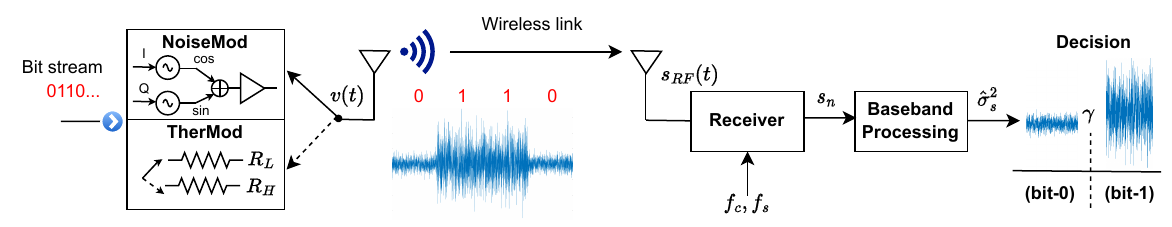}
		\vspace*{-0.3cm}\caption{Transceiver architecture of noise modulation. $v(t)$: thermal or externally-generated noise waveform, $s_{RF}(t)$: received signal scaled by a complex factor due to channel, $f_c/f_s$: carrier/sampling frequency.  For NoiseMod, $v(t)=\Re\left\lbrace u(t)e^{j2\pi f_c t } \right\rbrace $ with $u(t)$ being the complex baseband noise waveform.}\vspace*{-0.3cm}
		\label{fig:NoiseMod}
	\end{center}
\end{figure*} 

\IEEEPARstart{S}{ince}  the beginning of the digital communication era, amplitudes, phases, or frequencies of high-frequency carrier signals have been used to convey information. Index modulation (IM) challenges this status quo by using the indices of transmit entities to convey information \cite{Basar_2017}. However, in mainstream communication engineering, we always treat noise as an enemy to be dealt with and build our transceiver architectures based on this undeniable fact. \textit{What if there is gold in that garbage?} In other words, can we use noise, or random signals in general, to convey information?

The concept of \textit{wireless thermal noise modulation (TherMod)} was envisioned in the early work of \cite{Kish_2005} by two parabolic antennas driven by variable resistances (open, short, and 50 ohms). Later, the Kirchhoff-law-Johnson-noise scheme is proposed using thermal noises of two pairs of resistors to securely exchange two users' bits \cite{Kish_2006}. A few early designs have also been reported to convey information by random signal parameters (such as using the signal variance) \cite{Kramer_Bessai_2017,Basnayaka_Haas_2017,Kozlenko_Bosyi_2018,Anand_2019}. More recently, we have
provided a comprehensive background by using resistance IM to convey information. We obtained rich results on TherMod considering specific system parameters, such as information-carrying and disruptive noise powers \cite{Basar_2023}. However, a general framework for random signal modulation over wireless channels is still missing in the open literature, which is one of the main objectives of this letter.

The primary motivation of the \textit{noise modulation (NoiseMod)} concept is the exploitation of the noise variance to convey digital information. In general terms, we divide NoiseMod schemes into two categories: i) \textit{thermal noise modulation (Thermal-NoiseMod or shortly TherMod)} using background noise sources and ii) \textit{external noise modulation (External-NoiseMod or shortly NoiseMod)} using external and artificial noise sources. While TherMod might be suitable for potentially ultra-low/zero-power communication systems operating in very short ranges, general NoiseMod schemes, which are built on stochastic transceiver architectures, can be used in longer distances. Notably, simple electronic components such as resistors, capacitors, and switches might be used for TherMod to convey information bits. On the other hand, software-defined radio platforms or vector signal generators can be used to passband modulate and transmit complex baseband noise samples. As a result, in this letter, we investigate the error performance of TherMod schemes in the presence of line-of-sight (LOS)-dominated additive white Gaussian noise (AWGN) channels only; on the other hand, general NoiseMod system performance will be evaluated under fading channels. 

The rest of the letter is organized as follows. Section II provides a brief overview of the NoiseMod concept, followed by its analysis under AWGN and fading channels, respectively, in Sections III and IV. Sections V and VI are dedicated to two novel NoiseMod variants. Finally, Section VII provides numerical and computer simulation results along with conclusions in Section VIII.

\vspace*{-0.26cm}
\section{Concept of Noise Modulation}

The generic transceiver architecture of a NoiseMod system is given in Fig. 1. Here, depending on the requirements of the target system in terms of throughput, range, and power consumption, either artificially generated noise waveforms with adjustable transmit power or thermal (internal) noise sources, such as resistors, can be used. While TherMod can be implemented with simple circuitry, software-defined radios or signal generator platforms can be used for NoiseMod to in-phase/quadrature (IQ) modulate baseband noise samples and to obtain the passband noise waveform $v(t)$ at the expense of a higher power consumption. At the receiver of both designs, the received RF signal $s_{RF}(t)$ is downconverted to complex baseband samples $s_n$ by carefully selecting the carrier frequency $f_c$ and sampling frequency $f_s$. These baseband samples are further processed for bit detection, such as threshold-based detection shown in Fig. 1, by calculating the sample variance $\hat{\sigma}^2$ to extract the information embedded in noise variance.

The significant advantages of NoiseMod can be summarized as follows. First, the received samples can be regarded as stemming from an oversampled signal, allowing the flexibility of adjusting the number of noise samples per symbol $(N)$. This flexibility provides an exciting trade-off between error performance and bit rate. Furthermore, multiple noise samples of a single bit can be distributed in time at the transmitter to harvest the time diversity at the receiver without complex signal processing as in legacy systems. Embedding information in the noise variance also allows non-coherent detection at the receiver side without channel knowledge and with almost no performance degradation. Finally, using noise waveforms might be helpful in covert and unconditionally secure communications \cite{Mucchi_2022}.

In Fig. 2, we present the noise waveforms generated by the proposed three NoiseMod schemes in this letter. In simple terms, TherMod and NoiseMod schemes switch between low and high-variance noise waveforms for each bit interval, as in Fig. 2(a). Non-coherent (NC)-NoiseMod scheme switches from low-to-high or high-to-low at the middle of each bit duration, as shown in Fig. 2(b). Finally, the time diversity (TD)-NoiseMod scheme divides each bit duration into multiple slots, as in Fig. 2(c), and distributes the low or high-variance waveforms into multiple bit durations to harvest time diversity. In what follows, we provide the receiver architecture of each scheme along with their theoretical bit error probability (BEP) derivations.

\begin{figure}[!t]
\centering
		\includegraphics[width=1\columnwidth]{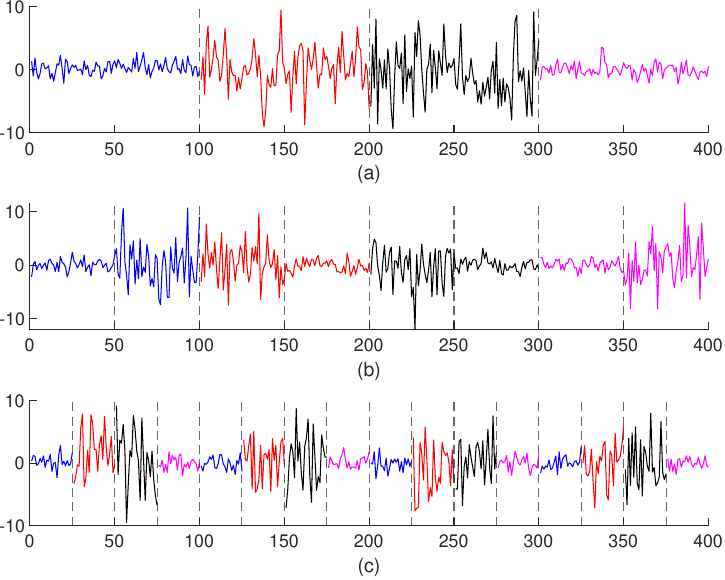}
		\vspace*{-0.3cm}\caption{Noise modulation waveforms for the bit string of $0110$ for $N=100$: (a) TherMod and NoiseMod, (b) NC-NoiseMod, and (c) TD-NoiseMod.}\vspace*{-0.3cm}
		\label{fig:Fig1}
\end{figure}

\section{Communication by Means of Thermal Noise: TherMod Revisited}
In this section, we revisit the TherMod scheme introduced in \cite{Basar_2023}, where the thermal noise generated by resistors is picked up by specific antennas. The terminology and notation used here will be taken as a baseline while developing novel NoiseMod schemes in later sections for fading channels.

Let us first review the error performance of the TherMod scheme. We express the $n$th received complex baseband sample as 
\begin{equation}\label{eq:1}
	s_n=r_n+w_n.
\end{equation}
Here, $r_n$ and $w_n$ stand for the information-carrying received noise sample and the AWGN sample, respectively, and the variance of $w_n$ is given by $\sigma_w^2$. We denote the variance of $r_n$ for bit-$0$ and bit-$1$ respectively by $\sigma_0^2$ and $\sigma_1^2$, where we assume $\sigma_1^2=\alpha \sigma_0^2$. In other words, $\alpha$ dictates the ratio of observed noise variances for two bit levels, and the variance of $r_n$ changes between low $(\sigma_0^2)$ and high $(\sigma_1^2)$ variance values for each signaling (bit) interval. In order to assess the error performance, we further define $\delta= \sigma_0^2 /\sigma_w^2$, which represents the ratio of variances of useful and distributive noise terms. Similar to signal-to-noise ratio in traditional digital communication systems, a higher $\delta$ means that the receiver is more robust against additive noise effects and a lower BEP can be obtained. In light of this discussion, $s_n$ follows complex Gaussian distribution and its variance is obtained as
\begin{equation}\label{eq:2}
 \sigma_s^2 = \begin{cases}
	\tilde{\sigma}_0^2 = \sigma_w^2 (1+\delta)  , \quad \,\,\,\, \text{for bit-}0\\
	\tilde{\sigma}_1^2 = \sigma_w^2 (1+\alpha\delta), \quad \text{for bit-}1.
\end{cases}
\end{equation}
In NoiseMod schemes, the task of the receiver is to extract the information embedded in the noise variance, as a result, we further process the received complex baseband samples for variance detection. Particularly, using $N$ samples for each bit duration of $T_b$ seconds, where a total of $f_s=N/T_b$ samples are processed per second, the variance of $s_n$ can be estimated as
 \begin{equation}\label{eq:3}
 	\hat{\sigma}_s^2=\frac{1}{N} \sum_{n=1}^{N} \left|s_n \right|^2. 
 \end{equation}
Considering the central limit theorem, $ \hat{\sigma}_s^2 $ follows Gaussian distribution, that is, $\mathcal{N}(\tilde{\sigma}_i^2,\tilde{\sigma}_i^4 /N)$ for $i\in \left\lbrace 0,1 \right\rbrace $. As shown in Fig. 1, the estimated variance is compared to a threshold to detect the transmitted bit $(b)$ as follows:
 \begin{equation} \label{eq:4}
 	\hat{b} = \begin{cases}
	0  , \quad \text{if}\quad\hat{\sigma}_s^2 < \gamma\\
	1, \quad \text{if} \quad\hat{\sigma}_s^2 > \gamma.
\end{cases}
 \end{equation}
Here, $\gamma$ stands for the threshold to be determined according to  $\alpha$ and $\delta$, and it can be scaled with respect to $\sigma_w^2$ as $\gamma= \chi \sigma_w^2$. Accordingly, the BEP of TherMod can be expressed as
\begin{align}
	P_b&=\frac{1}{2} \left[ P(\hat{\sigma}_s^2 >\gamma \left. \right| b=0 ) + P(\hat{\sigma}_s^2<\gamma \left. \right| b=1 ) \right] \nonumber \\
	&= \frac{1}{2} \left[ Q\left( \frac{\gamma-\tilde{\sigma}_0^2}{\sqrt{\tilde{\sigma}_0^4/N}} \right) + Q\left( \frac{\tilde{\sigma}_1^2-\gamma}{\sqrt{\tilde{\sigma}_1^4/N}} \right)  \right]. 
\end{align}
Substituting the values of $\gamma$, $\tilde{\sigma}_0^2$, and $\tilde{\sigma}_1^2$ and then canceling common $\sigma_w^2$ terms, we obtain
\begin{equation}\label{eq:6}
	P_b=\frac{1}{2}\left[ Q\left( \frac{\chi-(1+\delta)}{(1+\delta)\sqrt{1/N}} \right) + Q\left( \frac{(1+\alpha\delta)-\chi}{(1+\alpha\delta)\sqrt{1/N}} \right)  \right].
\end{equation}
Here, we note that $\chi$ is the scaled threshold value and it should be selected as $1+\delta < \chi < 1+\alpha\delta$. While minimization of \eqref{eq:6} for $\chi$ is a non-trivial task, we determine $\chi$ to ensure uniform BEP for bit-$0$ and bit-$1$, by equating the arguments of the two $Q$-functions of \eqref{eq:6}. For this purpose, we define $A=1+\delta$ and $B=1+\alpha\delta$, then solve $(\chi-A)/A=(B-\chi)/B$ for $\chi$ as 
\begin{equation}
	 \chi= \frac{2AB}{A+B}=\frac{2(1+\delta)(1+\alpha\delta)}{2+\delta(1+\alpha)}.
 \end{equation}
Accordingly, the BEP of TherMod is obtained as
\begin{equation}\label{eq:8}
	P_b=Q \left( \frac{\sqrt{N} (B-A)}{B+A} \right)=  Q\left(\frac{\sqrt{N}\delta(\alpha-1)}{2+\delta(\alpha+1)} \right). 
\end{equation}
As  discussed in \cite{Basar_2023}, $P_b$ given in \eqref{eq:8} reveals an exponentially decaying BEP for TherMod with respect to $N$ even for moderate $\delta$ and $\alpha$ values. Since a rich set of computer simulation results on the BEP of TherMod is presented in \cite{Basar_2023}, we shift our focus to general NoiseMod schemes.

\section{NoiseMod Under Fading Channels}
NoiseMod and TherMod waveforms are identical in nature and shown in Fig. 2(a). However, for the case of frequency non-selective fading, we assume that the noise samples generated by the NoiseMod transmitter are received from many paths, resulting in a fluctuating received signal variance from a bit to another. Accordingly, we rewrite \eqref{eq:1} as
\begin{equation}\label{eq:9}
	s_n= h r_n+w_n.
\end{equation}
Here $h$ stands for the small-scale fading coefficient and follows $\mathcal{CN}(0,1) $ distribution for Rayleigh fading. We further assume that $h$ remains constant for the transmission of a bit. In this case, $s_n$ still follows complex Gaussian distribution conditioned on $h$ and its variance is given as
\begin{equation}\label{eq:10}
	\sigma_s^2 = \begin{cases}
		s_0^2 = \sigma_w^2 (1+\left| h\right|^2 \delta)  , \quad \,\,\,\, \text{for bit-}0\\
		s_1^2 = \sigma_w^2 (1+\left| h\right|^2\alpha\delta), \quad \text{for bit-}1,
	\end{cases}
\end{equation}
where $\sigma_w^2,\delta$, and $\alpha$ are as defined in \eqref{eq:2}. In light of this information and considering a threshold-based detection as TherMod, that is, calculating the sample variance as $\hat{\sigma}_s^2=\frac{1}{N} \sum\nolimits_{n=1}^{N} \left|s_n \right|^2$, which follows the distribution of $\mathcal{N}(s_i^2,s_i^4 /N)$ for $i\in \left\lbrace 0,1 \right\rbrace $, and comparing it with a threshold as in \eqref{eq:4}, we obtain the conditional BEP as
\begin{equation}\label{eq:11}
	P_b=\frac{1}{2} \left[ Q\left( \frac{\gamma-s_0^2}{\sqrt{s_0^4/N}} \right) + Q\left( \frac{s_1^2-\gamma}{\sqrt{s_1^4/N}} \right)  \right].
\end{equation}
Here, $\gamma= \eta\sigma_w^2$ is the new threshold to be determined according to the wireless channel coefficient. Substituting the values of $\gamma$, $s_0^2$, and $s_1^2$ and then canceling common $\sigma_w^2$ terms, we obtain
\begin{equation}\label{eq:12}
	P_b\!=\!\frac{1}{2}\!\!\left[ Q\!\left(\! \frac{\eta-(1+\left| h\right|^2\delta)}{(1+\left| h\right|^2\delta)\sqrt{1/N}}\! \right) \!+\! Q\!\left(\! \frac{(1+\left| h\right|^2\alpha\delta)-\eta}{(1+\left| h\right|^2\alpha\delta)\sqrt{1/N}} \right)\!  \right]\!\!.
\end{equation}
Defining $C=1+\left| h\right|^2\delta$ and $D=1+\left| h\right|^2\alpha\delta$, and equating the arguments of the above $Q$-functions for uniform BEPs, we obtain the normalized threshold as
\begin{equation}\label{eq:13}
	\eta= \frac{2CD}{C+D}=\frac{2(1+\left| h\right|^2\delta)(1+\left| h\right|^2\alpha\delta)}{2+\left| h\right|^2 \delta(1+\alpha)}. 
\end{equation}
Substituting the value of $\eta$ of \eqref{eq:13} in \eqref{eq:12}, we obtain
\begin{equation}\label{eq:14}
	P_b=Q \left( \frac{\sqrt{N} (D-C)}{D+C} \right)=  Q\left(\frac{\sqrt{N}\left| h\right|^2\delta(\alpha-1)}{2+\left| h\right|^2\delta(\alpha+1)} \right). 
\end{equation}
Considering  the exponentially distributed probability density function (pdf) of $\left| h\right|^2$ as $f_{\left| h\right|^2}(u)= e^{-u}$ for $u\ge 0$ and the complicated nature of the conditional BEP expression of \eqref{eq:14}, we resort to numerical integration to obtain the unconditional BEP as follows:
\begin{align}\label{eq:15}
	\bar{P_b}&= \mathrm{E}\left[  Q\left(\frac{\sqrt{N}\left| h\right|^2\delta(\alpha-1)}{2+\left| h\right|^2\delta(\alpha+1)} \right)\right] \nonumber \\
	&= \int_{0}^{\infty} Q\left(\frac{\sqrt{N}u\delta(\alpha-1)}{2+u\delta(\alpha+1)} \right) \exp(-u) du. 
\end{align}
We note that an alternative solution can be obtained by trying to fit the distribution of argument of the $Q$-function in \eqref{eq:14} with numerical methods and then using moment generation function approach to derive the unconditional BEP, however, we leave this non-trivial task to interested readers. For other fading distributions, the pdf of $\left| h\right|^2$ in \eqref{eq:15} can be updated.

While a closed-form expression is not available for $\bar{P_b}$, our numerical integration results, shown in Section VII, have revealed that $\bar{P_b} \propto 1/(N \delta)$ for NoiseMod. This shows a linear decrease in BEP with respect to $N$ and $\delta$, as opposed to exponential decrease in BEP of TherMod, however, increasing $N$ can still improve the BEP. On the other hand, this poor BEP performance under fading has led us to develop diversity schemes, which will be discussed in Section VI.

Threshold-based detector of NoiseMod adjusts $\eta$ in \eqref{eq:13} considering the instantaneous value of $\left| h\right|^2$, as a result, its requires channel knowledge. This requirement has led us to explore alternative designs that can operate without channel knowledge, as discussed in the next section.

\section{Non-Coherent NoiseMod}

NC-NoiseMod builds on the idea of detecting the transmitted bit by processing the received noise samples without channel knowledge. For this purpose, taking inspiration from Manchester line coding scheme, we switch between low and high variance values at the middle of each bit period. As shown in Fig. 2(b), for bit-$0$, for the first half of the bit period, we generate a noise waveform of low variance and switch to the high variance samples in the second half, while doing the reverse for bit-$1$. 

The detector of NC-NoiseMod operates as follows.  Processing the same samples of \eqref{eq:9}, we calculate two sample variances corresponding to first and second halves of the bit duration:
 \begin{equation}\label{eq:16}
	\hat{\sigma}_{s,1}^2=\frac{1}{N/2} \sum_{n=1}^{N/2} \left|s_n \right|^2, \quad \hat{\sigma}_{s,2}^2=\frac{1}{N/2} \sum_{n=N/2+1}^{N} \!\!\!\!\left|s_n \right|^2.
\end{equation}
Then we simply detect the transmitted bit without any channel knowledge as
 \begin{equation} \label{eq:17}
	\hat{b} = \begin{cases}
		0  , \quad \text{if}\quad\hat{\sigma}_{s,1}^2< \hat{\sigma}_{s,2}^2\\
		1, \quad \text{if} \quad\hat{\sigma}_{s,1}^2> \hat{\sigma}_{s,2}^2.
	\end{cases}
\end{equation}

The conditional BEP of NC-NoiseMod is expressed as 
\begin{equation}
		P_b=\frac{1}{2} \left[ P(\hat{\sigma}_{s,1}^2> \hat{\sigma}_{s,2}^2 \left. \right| b=0 ) + P(\hat{\sigma}_{s,2}^2>\hat{\sigma}_{s,1}^2 \left. \right| b=1 ) \right].
\end{equation}
Here, we note that $\hat{\sigma}_{s,1}^2 \sim \mathcal{N}(s_0^2,s_0^4/(N/2))$ and $\hat{\sigma}_{s,2}^2 \sim \mathcal{N}(s_1^2,s_1^4/(N/2))$ for bit-$0$ and $\hat{\sigma}_{s,1}^2 \sim \mathcal{N}(s_1^2,s_1^4/(N/2))$ and $\hat{\sigma}_{s,2}^2 \sim \mathcal{N}(s_0^2,s_0^4/(N/2))$ for bit-$1$, and $s_0^2$ and $s_1^2$ are as defined in \eqref{eq:10}. Defining $D_1=\hat{\sigma}_{s,1}^2- \hat{\sigma}_{s,2}^2 $ and $D_2=\hat{\sigma}_{s,2}^2- \hat{\sigma}_{s,1}^2 $, we obtain
\begin{equation}\label{eq:19}
	P_b=\frac{1}{2} \left[ P(D_1>0 \left. \right| b=0 ) + P(D_2>0 \left. \right| b=1 ) \right].
\end{equation}
Considering the conditional distributions of $D_1$ and $D_2$, which are identical and given as $\mathcal{N}\left( s_0^2-s_1^2 ,\frac{s_0^4+s_1^4}{N/2}\right) $, we obtain
\begin{equation}\label{eq:20}
	P_b=Q\left(\frac{s_1^2-s_0^2}{\sqrt{(s_1^4+s_2^4) /(N/2)}} \right).
\end{equation}
Substituting the values of $s_0^2 = \sigma_w^2 (1+\left| h\right|^2 \delta)$ and $s_1^2 = \sigma_w^2 (1+\left| h\right|^2 \alpha\delta)$ in \eqref{eq:20} and after simple manipulations, the conditional BEP is expressed as
 \begin{equation}\label{eq:21}
 	P_b= Q\left(\frac{\sqrt{N/2} \left| h\right|^2 \delta(\alpha-1)}{\sqrt{(1+\left| h\right|^2\delta\alpha)^2 + (1+\left| h\right|^2\delta)^2} } \right). 
 \end{equation}
Similar to the NoiseMod scheme's unconditional BEP calculation of \eqref{eq:15}, numerical integration can be used to take the expectation of \eqref{eq:21} over the fading distribution. 

As seen from \eqref{eq:21}, the conditional BEP of NC-NoiseMod is still a function of $N$, $\delta$, and $\alpha$, however, it is not easy to catch useful insights by comparing it with NoiseMod's conditional BEP given by \eqref{eq:14}. The relative error performance difference of NC-NoiseMod compared to the NoiseMod scheme will be evaluated by computer simulations. 

\section{Time-Diversity NoiseMod}
The concept of TD-NoiseMod has been put forward considering the following two facts. First, the BEP performance of NoiseMod schemes is not satisfactory under fading channels. Second, since multiple samples are taken for each bit, distributing these samples over time might be a remedy to circumvent small-scale fading effects as long as different groups of noise samples belonging to the same bit experience independent fading channels.

As shown in Fig. 2(c), in TD-NoiseMod, the given bit duration of $N$ samples are divided into $I$ slots, where $I$ also determines the time spreading ratio. That is, we spread the variance information by generating noise samples belonging to the same bit for $I$ different slots. For the example of Fig. 2(c), $I=4$ and $N/4$ samples belonging to four different slots are used for each bit.

Considering the operation of TD-NoiseMod and the noise samples of $\eqref{eq:9}$ for a specific bit in $I$ time distributed slots, the variance of $s_n$ becomes
\begin{equation}
	\sigma_s^2 = \begin{cases}
		s_0^2 = \frac{1}{I}\sum\nolimits_{i=1}^{I}\sigma_w^2 (1+\left| h_i\right|^2 \delta)  , \quad \,\,\,\, \text{for bit-}0\\
			s_1^2 = \frac{1}{I}\sum\nolimits_{i=1}^{I}\sigma_w^2 (1+\left| h_i\right|^2 \alpha\delta) , \quad \text{for bit-}1.
	\end{cases}
\end{equation} 
Here $\delta$ and $\alpha$ are as defined in \eqref{eq:2} and \eqref{eq:10}. $h_i\sim \mathcal{CN}(0,1)$ stands for the small-scale fading coefficient belonging to the $I$th bit duration and assumed to be independent from one bit duration to another. We note that TD-NoiseMod can still be operational when $h_i$ changes more slowly compared to the bit duration by distributing noise samples of the same bit further apart in time. 

Defining a new threshold as $\gamma=\kappa \sigma_w^2$, applying the threshold-based detection of \eqref{eq:4}, and considering the conditional BEP expression of \eqref{eq:11}, we obtain
\begin{equation}\label{eq:23}
	P_b\!=\!\frac{1}{2}\!\!\left[ Q\!\left(\! \frac{\kappa-E}{E\sqrt{1/N}}\! \right) \!+\! Q\!\left(\! \frac{F-\kappa}{F\sqrt{1/N}} \right)\!  \right]\!\!,
\end{equation}
where $E=\frac{1}{I}\sum\nolimits_{i=1}^{I}\sigma_w^2 (1+\left| h_i\right|^2 \delta)$ and $F=\frac{1}{I}\sum\nolimits_{i=1}^{I}\sigma_w^2 (1+\left| h_i\right|^2 \alpha\delta)$. In light of this, selecting $\kappa=2EF/(E+F)$ to have uniform BEPs and substituting $\kappa$ in \eqref{eq:23}, we obtain 
\begin{equation}\label{eq:24}
	P_b=Q \left( \frac{\sqrt{N} (F-E)}{F+E} \right)=  Q\left(\frac{\sqrt{N}\delta(\alpha-1)g}{2+\delta(\alpha+1)g} \right). 
\end{equation}
where $g=\sum_{i=1}^{I}\left| h_i\right|^2$. It is worth noting that selecting $I=1$ in \eqref{eq:24}, one obtains the conditional BEP of NoiseMod given in \eqref{eq:14}. Here, unconditional BEP of TD-NoiseMod can be obtained by numerical integration again considering the chi-square distribution of $g$ with $2I$ degrees of freedom. Specifically, $f_g(u)=2(2u)^{I-1}e^{-u}/(2^I \Gamma(I))$, where $\Gamma(\cdot)$ is the Gamma function, and the unconditional PEP of TD-NoiseMod is given by
\begin{equation}\label{eq:25}
\bar{P_b}=\int_{0}^{\infty}  Q\left(\frac{\sqrt{N}u\delta(\alpha-1)}{2+u\delta(\alpha+1)} \right)  f_g(u) du.
\end{equation}
The big question after the derivation of unconditional BEP of TD-NoiseMod is ``Does TD-NoiseMod provide a diversity order of $I$?" Unfortunately, we do not have an analytical and closed-form result to answer this question since we can numerically evaluate \eqref{eq:25} only. Nevertheless, our numerical analysis from Section VII shows that $\bar{P_b} \propto 10^{1-I} /(N \delta^I) $ for $I\in \left\lbrace 2,3,4,5\right\rbrace $, which proves that a diversity order of $I$ is attainable in terms of $\delta$.

\section{Numerical Results and Simulations}
In this section, we provide numerical and computer simulation results for TherMod, NoiseMod, NC-NoiseMod, and TD-NoiseMod. In all simulations, we consider a noise variances ratio of $\alpha=10$ and vary the number of noise samples per bit $(N)$ or the ratio of useful/disruptive noise variances $(\delta)$.

In Fig. 3, we provide the results of our numerical integrations to assess the BEP of various NoiseMod schemes by varying $\delta$ between $-10$ and $+10$ dB for the fixed value of $N=120$. For reference, the BEP of TherMod, which operates under pure AWGN channels, is also shown in this figure. As shown in Fig. 3, NoiseMod does not provide diversity and its BEP is proportional to $1/(N \delta)$. More interestingly, NoiseMod and NC-NoiseMod schemes provide almost the same BEP, while the latter does not require channel knowledge for threshold selection in bit detection. Considering this fact, NC-NoiseMod can be more favorable for simple detection at the expense of more frequent low-high noise variance switching at the transmitter. On the other hand, a diversity order of $I$ is attainable for TD-NoiseMod schemes considering the fact that their numerical BEP approximately fits the reference curves of $10^{1-I} /(N \delta^I)$ for the considered $N$ and $I$ values. 

\begin{figure}[!t]
	\centering
	\includegraphics[width=0.78\columnwidth]{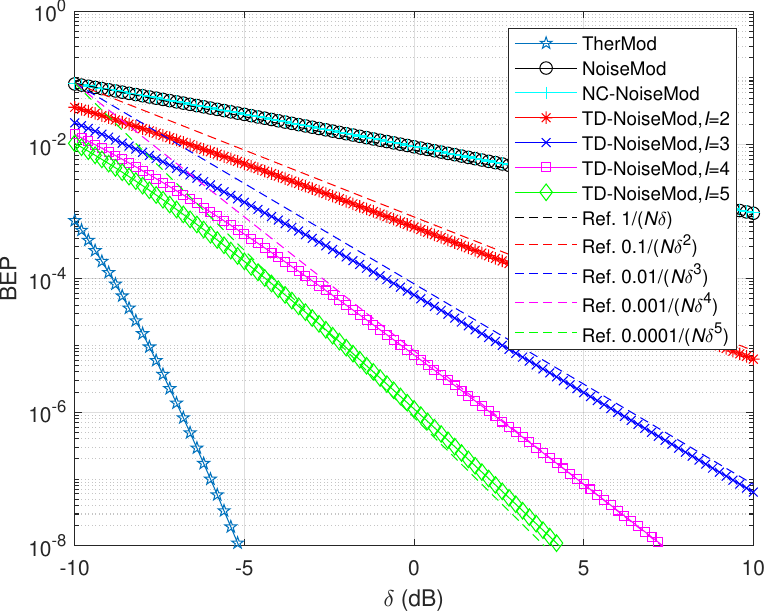}
	\vspace*{-0.3cm}\caption{Numerical BEP results for various NoiseMod schemes.}\vspace*{-0.3cm}
	\label{fig:Fig2}
\end{figure} 

In Fig. 4, we compare the theoretical BEP and simulation bit error rate (BER) results for NoiseMod and TD-NoiseMod schemes. Here, we assume $I=2$ for TD-NoiseMod and two cases: $N=100$ and $N=150$. As seen from Fig. 4, our theoretical BEP calculations based on numerical integration perfectly matches Monte Carlo simulation results for all considered cases. It is worth noting that increasing $N$ from $100$ to $150$ improves the BER performance by approximately $1$ dB for both schemes and the diversity advantage of TD-NoiseMod is again clearly visible. During the production of Fig. 4, we observed that the values for $N<100$, particularly for TD-NoiseMod, can exhibit inconsistent BER results even with extremely high $\delta$ values. This can be explained by the fact that all our derivations are based on CLT with a Gaussian distributed sample variance $(\hat{\sigma}_s^2)$ and and an insufficient number of noise samples can cause statistical decision errors regardless of $\delta$. We also note that increasing $N $ might be helpful for overcoming practical issues that affect the distinguishability of low and high variance waveforms.

Finally, in Fig. 5, we investigate the BER performance of various NoiseMod schemes with increasing $N$ values. This figure shows the delicate dependency on $\delta$ values, where for smaller $\delta$, even high $N$ values cannot escape from error floors. As seen from Fig. 5, increasing $\delta$ eventually improves the BER performance. While NoiseMod and NC-NoiseMod achieves identical BER as noted earlier, TD-NoiseMod with $I=2$ shines with its superior performance.

\begin{figure}[!t]
	\centering
	\includegraphics[width=0.78\columnwidth]{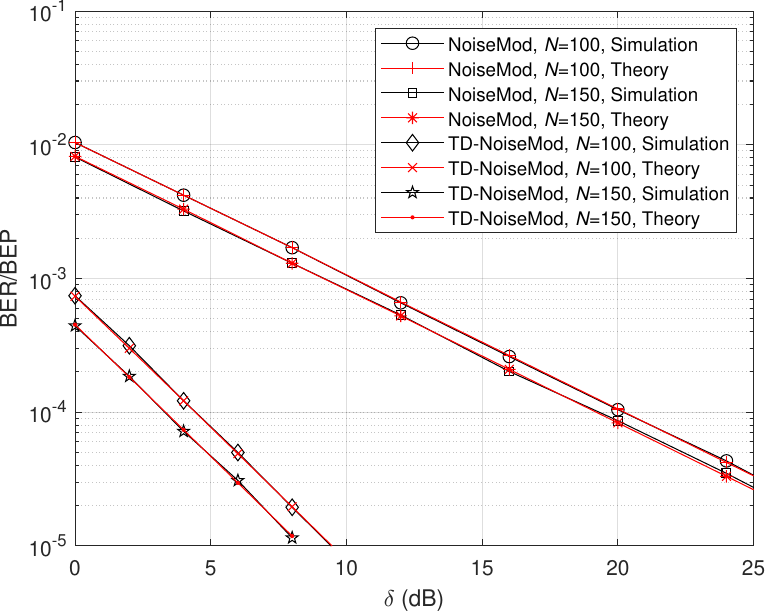}
	\vspace*{-0.3cm}\caption{Comparison of theoretical and simulation results for NoiseMod and TD-NoiseMod for $I=2$.}\vspace*{-0.2cm}
	\label{fig:Fig3}
\end{figure}

\begin{figure}[!t]
	\centering
	\includegraphics[width=0.9\columnwidth]{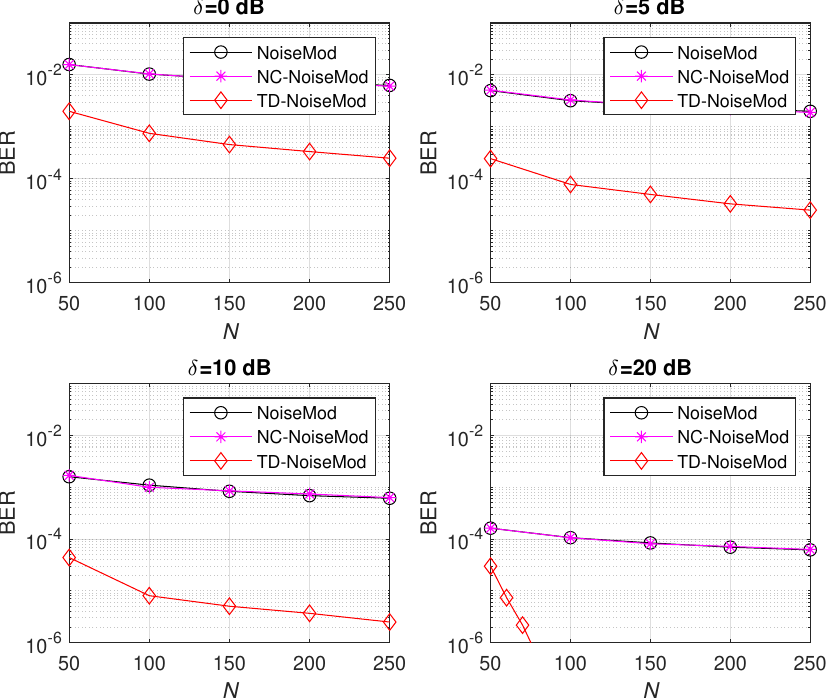}
	\vspace*{-0.3cm}\caption{BER performance of various NoiseMod schemes for increasing $N$.}\vspace*{-0.3cm}
	\label{fig:Fig4}
\end{figure} 

\section{Conclusions}
In this letter, we have laid out the theoretical fundamentals of communication by means of noise-alike signals. Notably, the term NoiseMod has been coined to cover the concept of digital data transmission with internally or externally generated noise signals. Hopefully, this letter will pave the way for generalized and more sophisticated designs with practical testbeds for NoiseMod schemes in future studies. Evaluation of NoiseMod for secure and covert communications is also an exciting research direction. While an initial practical experiment verified our theoretical assessment and the distinguishability of noise waveforms with different variances, interesting practical issues are left to our future studies.

\vspace*{-0.26cm}
\bibliographystyle{IEEEtran}
\bibliography{IEEEabrv,bib_2023}

\end{document}